\documentclass[runningheads]{lncse}

\usepackage{amsmath}
\usepackage{amsfonts}
\usepackage{epsfig}
\usepackage{graphics} 
\usepackage{tikz, pgfplots}
\usetikzlibrary{plotmarks}

\usepackage{stmaryrd}
\usepackage{import, color}

\begin{document}

\title{Reactivation of Fractures in Subsurface Reservoirs -- a Numerical
  Approach using a Static-Dynamic Friction Model}

% your contribution title if the original one is too long use an abbreviated
% title (for running head):
\titlerunning{Reactivation of Fractures in Subsurface Reservoirs}

\author{ Runar L. Berge\inst{1} \and Inga Berre \inst{1} \and Eirik Keilegavlen
  \inst{1}}

% Use \authorrunning{Short Title} for an abbreviated version of the author list
% (for running head):
\authorrunning{Berge et al.}

\institute{ University of Bergen, Department of Mathematics, {\tt
    Runar.Berge@uib.no} }

\maketitle

\begin{abstract}
  Fluid-induced slip of fractures is characterized by strong multiphysics
  couplings. Three physical processes are considered: Flow, rock deformation
  and fracture deformation. The fractures are represented as lower-dimensional
  objects embedded in a three-dimensional domain. Fluid is modeled as slightly
  compressible, and flow in both fractures and matrix is accounted for. The
  deformation of rock is inherently different from the deformation of
  fractures; thus, two different models are needed to describe the mechanical
  deformation of the rock. The medium surrounding the fractures is modeled as a
  linear elastic material, while the slip of fractures is modeled as a contact
  problem, governed by a static-dynamic friction model. We present an iterative
  scheme for solving the non-linear set of equations that arise from the
  models, and suggest how the step parameter in this scheme should depend on
  the shear modulus and mesh size.
\end{abstract}

\section{Introduction}
\label{berge_mini_2:sec:introduction}

Slip or reactivation of fractures and faults in the earth's crust can be caused
by natural changes in the stress field in the rock, but concerns also arise
related to human induced seismicity. Over 700 cases of seismicity related to
human activities have been reported, caused by a variety of different
activities~\cite{berge_mini_2:foulger2017global}. In this paper, we focus on
fracture reactivation due to injection of fluids in underground reservoirs,
which concerns applications such as CO$_2$ storage, enhanced oil recovery and
production of geothermal energy. The seismic events related to these types of
activities are usually not noticeable except by very sensitive equipment, but
events of magnitude up to M3.5 have been
recorded~\cite{berge_mini_2:zang2014analysis}.

Reactivation of fractures is a strongly coupled problem involving disparate
physical processes, including fluid flow, deformation of rock surrounding the
fractures and the plastic deformation of fractures. In this work we will couple
models for each of these sub-problems to simulate fracture reactivation and
corresponding aperture changes. One of the main challenges of modeling
fractures in the subsurface is the absence of scale separation between
fractures, which can range from sub-resolution micro-fractures to fractures and
faults spanning the whole reservoir. In the model presented here, small scale
fractures are assumed to be up-scaled into effective matrix parameters, such as
permeability, % which can be done in a various of ways (e.g., see
% \cite{berge_mini_2:mourzenko2011permeability,saelvik2014anisotropic}).
while the large macroscopic fractures are explicitly resolved. To reduce the
computational cost further, fractures are represented as lower-dimensional
domains with an associated aperture. We apply this mixed dimensional hybrid
approach to model fluid flow following the ideas
of~\cite{berge_mini_2:dietrich2005flow}.

% Induced seismicity is not always negative, it is even wanted in some
% cases. In enhanced geothermal systems one tries to reactivate fractures in
% order to increase the permeability of the fracture network
% \cite{berge_mini_2:}. The processes leading to seismicity are highly coupled,
% varies on multiple time and space scales, and our understanding of them is in
% many aspect limited.

The fluid pressure will act as a trigger mechanism for fracture reactivation,
which is governed by a Mohr-Coulomb criterion: A fracture slips when the shear
traction equals the coefficient of friction times the effective normal
traction. This bound on shear traction introduces a non-linear inequality
constraint to the system of equations, and various techniques have been
employed to handle this
constraint~\cite{berge_mini_2:aagaard2013domain,berge_mini_2:kikuchi1988contact}.
% presents a method that iterate between fixing shear stress and normal stress,
% while \cite{berge_mini_2:aagaard2013domain} presents an iterative scheme
% using finite elements and Lagrangian multipliers.
We use a static-dynamic friction model where the coefficient of friction drops
instantaneous when a fracture slips. Results from this model should be
interpreted with caution, as it falls into the category of "inherently
discrete" models due to this discontinuous
drop~\cite{berge_mini_2:rice1993spatio‐temporal}. Nevertheless, it has been
shown to give feasible results in modeling of fracture reactivation as a
consequence of fluid injection at elevated
pressures~\cite{berge_mini_2:mcClure2011investigation}. Finding the regions of
slip on a fracture is one of the challenges of this model, due to the
discontinuous change in the coefficient of friction between stick and slip
regions. We will give a mathematical formulation of the friction problem, and
then present a simple solution strategy following the ideas Ucar et
al.~\cite{berge_mini_2:ucar2017three}, who use an idea of excess shear stress
to estimate fracture slip based on the shear and normal stress. The slip length
that each time step was approximated based on how much the Mohr-Coulomb
criterion is violated, multiplied a ``fracture stiffness'' parameter. In the
current work, we improve this approach by suggesting how this stiffness
parameter should depend on the shear modulus and mesh size.

% numerical solutions of friction models for the reactivation of pre-existing
% fracture networks and the treatment of the non-linearities introduced by the
% slip. We use an empirically validated model of friction.

% In this paper we will limit our self to induced seismicity due to injection
% of fluids.  More specifically, we will investigate Enhanced Geothermal
% Reservoirs (EGS). In EGS, seismicity is induced on purpose in order to
% increase the permeability of the reservoir. In many cases this is necessary
% to achieve commercial flow rates. By doing so, the permeability can increase
% by orders of magnitude, which is considered the key to unlock geothermal
% resources around the world. In EGS water at high pressure is injected into
% the fracture network in order to make pre-existing fractures slip. The
% fracture aperture can then increase due to asperities along the fracture
% faces, which will increase the permeability of the fractures. However, if the
% induced seismicity grows to large it can lead to bad public perception of
% geothermal project or even the closure of costly projects. It is therefore
% important to create good models of this phenomena in order to understand the
% mechanisms that causes large seismic events.

\section{Governing Equations}
\label{berge_mini_2:sec:model}
\begin{figure}
  \centering
  \begin{minipage}{0.3\textwidth}
    \centering \def\svgwidth{\textwidth}%
    \begingroup%
    \large%
    \makeatletter%
    \providecommand\color[2][]{%
      \errmessage{(Inkscape) Color is used for the text in Inkscape, but the
        package 'color.sty' is not loaded}%
      \renewcommand\color[2][]{}%
    }%
    \providecommand\transparent[1]{%
      \errmessage{(Inkscape) Transparency is used (non-zero) for the text in
        Inkscape, but the package 'transparent.sty' is not loaded}%
      \renewcommand\transparent[1]{}%
    }%
    \providecommand\rotatebox[2]{#2}%
    \ifx\svgwidth\undefined%
    \setlength{\unitlength}{870bp}%
    \ifx\svgscale\undefined%
    \relax%
    \else%
    \setlength{\unitlength}{\unitlength * \real{\svgscale}}%
    \fi%
    \else%
    \setlength{\unitlength}{\svgwidth}%
    \fi%
    \global\let\svgwidth\undefined%
    \global\let\svgscale\undefined%
    \makeatother%
    \begin{picture}(1,1.02298851)%
      \put(0,0){\includegraphics[width=\unitlength]{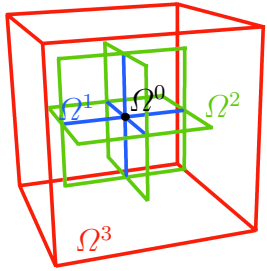}}%
      \put(0.34447751,0.07327006){\color[rgb]{0.97254902,0.11764706,0.03921569}\makebox(0,0)[b]{\smash{$\Omega^3$}}}%
      \put(0.7585605,0.58041368){\color[rgb]{0.36470588,0.79607843,0.03529412}\makebox(0,0)[lb]{\smash{$\Omega^2$}}}%
      \put(0.27551752,0.57035701){\color[rgb]{0.12941176,0.36078431,1}\makebox(0,0)[b]{\smash{$\Omega^1$}}}%
      \put(0.47410026,0.60483707){\color[rgb]{0,0,0}\makebox(0,0)[lb]{\smash{$\Omega^0$}}}%
    \end{picture}%
    \endgroup%
  \end{minipage}
  \hspace{0.05\textwidth}
  \begin{minipage}[b]{0.5\textwidth}
    \caption{ A mixed dimensional domain.  The red cube represents the 3D
      domain, the green squares represent the 2D domains, the blue lines
      represent 1D domains, and the black dot the 0D
      domain \label{berge_mini_2:fig:mixedDim}}
  \end{minipage}
\end{figure}

We consider three processes in the subsurface; fluid flow, rock deformation,
and fracture deformation. Each of the problems will be defined in different
domains of different dimensions, and the domains and processes are coupled
together.  As mentioned above, fractures are represented as lower dimensional
objects in the reservoir. The intersection of two fractures is,
correspondingly, represented by a line-segment. The reservoir is in this way
divided into domains~$\Omega^d$ of different dimension $d$; the rock matrix
$\Omega^3$, fractures $\Omega^2$, intersection of fractures $\Omega^1$, and the
intersection points of these lines $\Omega^0$, as shown in
Fig.~\ref{berge_mini_2:fig:mixedDim}.

\subsection*{Flow}

We will assume that the fluid flow follows Darcy's law in all domains, and
extend the approach of Boon et al.~\cite{berge_mini_2:boon2016robust} to
compressible flow. For a detailed explanation of notation we refer the reader
to the same work. The mixed-dimensional formulation for the conservation of
mass is given by
\begin{equation}\label{berge_mini_2:eq:flow_equation}
  \phi c_p\frac{\partial p}{\partial t}
  - \nabla \cdot \mathcal K\nabla p + \mathcal T
  = q
  \qquad \text{in}\ \{\Omega^d\}_{d=0,\ldots,3},
\end{equation}
where $\phi$ is the porosity, $c_p$ the fluid compressibility, $\mathcal K$~the
effective permeability (taking into account the scaling with fracture aperture
and viscosity), $p$ the fluid pressure, $\mathcal T$ the transfer term between
domains of different dimensions, and $q$ the source and sink term.

\subsection*{Rock Deformation}

The rock is modeled as an elastic medium, and assumed to always be in
quasi-static equilibrium. The conservation of momentum can be written as
\begin{equation}\label{berge_mini_2:eq:elastic_equation}
  \nabla \cdot \vec \sigma = 0,\quad
  \vec \sigma = G (\nabla \vec u + (\nabla \vec u)^\top) + \lambda
  \vec I\nabla\cdot \vec u \quad \text{in}\ \Omega^3,
\end{equation}
where as for the flow, we have neglected the gravitational term. The
variable~$\vec \sigma$ is the stress tensor, $G$ and $\lambda$ are the Lam\'e
parameters, $\vec I$ the identity matrix, and $\vec u$ the displacement
vector. Because a fracture will have exactly two interfaces with $\Omega^3$, we
adopt the notation that $\Gamma^+$ defines one side, while $\Gamma^-$ the
other.

We will neglect any elastic effects of contacting asperities of the fracture
surfaces and potential gouge filling the fractures, thus, the traction on the
two fracture boundaries must balance
\begin{equation}\label{berge_mini_2:eq:force_balance}
  \vec T^+ + \vec T^- = 0,
\end{equation}
where the traction is defined as $\vec T = \vec \sigma \cdot \vec n$, with
$\vec n$ being the normal vector of the surface.

In addition to boundary conditions on the reservoir domain, we need a total of
six equations defined on the fracture interfaces since we have three degrees of
freedom on each fracture side. The first three are obtained from the force
balance Equation~\eqref{berge_mini_2:eq:force_balance} and the last three
equations are defined by the coupling to the fracture deformation model, which
governs the fracture response to rock stress and strain.

% but it is worth noting that the coupling between the models happens on the
% interface between the fractures and the rock.

% Below we will present a solution strategy which defines periodic type
% boundary conditions across the fractures; the difference
% $\vec u^+ - \vec u^-$ on the positive and negative side of the fracture is
% assumed given from the fracture deformation model.

\subsection*{Fracture Deformation}
% which makes both the mathematical formulation and solution to this type of
% contact problem hard.

For two surfaces in contact, the magnitude of the shear traction is bounded by
the coefficient of friction times the effective normal traction
\begin{equation}\label{berge_mini_2:eq:mohr-coulomb}
  \lvert \vec T_s\rvert \leq \mu (T_n - p),
\end{equation}
where $\vec T_s$ is the shear traction, $\mu$ the coefficient of friction, and
$T_n$ the normal traction. We use a static-dynamic friction model, where the
coefficient takes one value when the fracture is not slipping $\mu = \mu_s$
(static friction), and another, lower value when it is slipping $\mu = \mu_d$
(dynamic friction).
%
% More sophisticated models have been
% used~\cite{berge_mini_2:dieterich1972time,dieterich1079modeling,ruina1983slip},
% however, the static-dynamic friction model has been shown to give

We introduce the slip distance $\vec d$ which defines the relative fracture
displacement
\begin{equation*}
  \vec d = \vec u^+ - \vec u^-,
\end{equation*}
where $\vec u^+$ and $\vec u^-$ are the displacements on the positive and
negative side of a fracture. When a fracture slips, the aperture can increase
due to asperities on the fracture surfaces, and we approximate this increase by
letting the aperture depend linearly on slip distance.

According to the Mohr-Coulomb criterion, slip on a fracture is triggered when
Inequality~\eqref{berge_mini_2:eq:mohr-coulomb} reaches equality, using the
static coefficient of friction. We define $\Gamma_s$ to be the part of the
fracture that is slipping, and enforce
Inequality~\eqref{berge_mini_2:eq:mohr-coulomb} as an equality in this domain,
using the dynamic coefficient of friction. Further, when a fracture is slipping
it should slip in the direction of shear traction. 

To sum up, the friction condition on the fractures can be formulated as
% \begin{subequations}\label{berge_mini_2:eq:frac_equation}
%   \begin{align}
%     &\vec d(\lvert \vec T_s\rvert - \mu_d(T_n - p)) = 0 \ &&\quad \vec x \in \Omega^2\\
%     & \lvert \vec T_s\rvert \leq \mu (T_n - p) && \quad \vec x \in \Omega^2\label{berge_mini_2:eq:frac_equation_slip}\\
%     &\Gamma_s = f(\vec u),&&\label{berge_mini_2:eq:frac_equation_domain}
%   \end{align}
% \end{subequations}

\begin{subequations}\label{berge_mini_2:eq:frac_equation}
  \begin{align}
    &\lvert \vec T_s\rvert < \mu_s(T_n - p)
    && \vec x\in\Omega^2\setminus\Gamma_s\label{berge_mini_2:eq:frac_equation_mohr}\\
    &\vec d = 0
    && \vec x \in \Omega^2\setminus\Gamma_s\\
    &\lvert \vec T_s\rvert - \mu_d(T_n - p) = 0
    && \vec x \in\Gamma_s\label{berge_mini_2:eq:frac_equation_slip}\\
    &\vec d = \chi \vec T_s,\ \chi > 0
    && \vec x \in \Gamma_s\label{berge_mini_2:eq:frac_equation_step}\\
    &\Gamma_s = \{\vec x\in \Omega^2: \vec d \neq 0\},&&\label{berge_mini_2:eq:frac_equation_domain}
  \end{align}
\end{subequations}
where we have included the definition of the slip region to stress the fact
that $\Gamma_s$ is not known, but one of the unknowns in this system.

\section{Spatial Discretization}
Both the flow equation and the elasticity problem are discretized using finite
volume schemes. The flow equation is discretized using the two-point flux
approximation, while the elasticity is discretized using the multi-point stress
approximation~\cite{berge_mini_2:keilegavlen2017finite,berge_mini_2:ucar2016finite}. The
implementation has been done in PorePy, which is an open source Python code for
simulating fractured and deformable porous
media~\cite{berge_mini_2:keilegavlen2017porepy}.

\section{Solution Strategy}
\label{berge_mini_2:sec:solution-strategy}

We will use an iterative scheme to solve the set of
Equations~\eqref{berge_mini_2:eq:flow_equation},
\eqref{berge_mini_2:eq:elastic_equation}
and~\eqref{berge_mini_2:eq:frac_equation}. At each step in the scheme we will
first estimate the slip region $\Gamma_s$ and then solve the equations using a
linearized approximation of
Equation~\eqref{berge_mini_2:eq:frac_equation_slip}.

We start by solving Equation~\eqref{berge_mini_2:eq:flow_equation} using
backward Euler. To solve \eqref{berge_mini_2:eq:elastic_equation}
and~\eqref{berge_mini_2:eq:frac_equation}, we first assume that the fractures
do not slip at the current time step, $\Gamma_s = \emptyset$, thus, the current
slip vector $\vec d^{k+1}$ must equal the slip vector at the previous time step
$\vec d^k$. Together with the stress condition, $\vec T^+ =- \vec T^-$, the
number of unknowns equals the number of equations and we have a closed
system. After we obtain the displacement $\vec u^{k+1}$, we can calculate the
stress, and from this traction, on the fractures from
Equation~\eqref{berge_mini_2:eq:elastic_equation}. We check if the Mohr-Coulomb
criterion~\eqref{berge_mini_2:eq:frac_equation_mohr} is satisfied; if it is,
our assumption that we have no slip is good, and we continue to the next time
step. If the condition is violated, we update the slip region $\Gamma_s$ by
including all faces that violate the condition. We define the ``excess shear
stress'' to be the residual of
Equation~\eqref{berge_mini_2:eq:frac_equation_slip},
\begin{equation*}
  T_e = \lvert \vec T_s\rvert - \mu_d(T_n - p) \quad \vec x\in \Gamma_s,
\end{equation*}
where we use the dynamic coefficient of friction. From
Equation~\eqref{berge_mini_2:eq:frac_equation_step}, we know that the slip
should be in the direction of shear stress, but the variable $\chi$ is
unknown. We make a simple estimate of this distance by assuming that all
degrees of freedom are fixed except on a single face $\mathcal F_i$ on the
fracture. The displacement gradient $\nabla \vec u_i$ at this face then grows
linearly with slip distance, and from
Equation~\eqref{berge_mini_2:eq:elastic_equation} we obtain that the
corresponding change in shear traction grows as
\begin{equation*}
  \Delta\vec T_s \sim \frac{G}{\lvert \mathcal F_i \rvert^{1/2}}\vec d_i,
\end{equation*}
where $\lvert \mathcal F_i\rvert$ is the area of the face. For each face that
is slipping, we therefore set our new guess for the slip vector to
\begin{equation*}
  \vec d_i^{k+1} = \vec d_i^k + \frac{\gamma \lvert \mathcal
    F_i\rvert^{1/2}}{G}T_e\vec \tau,
\end{equation*}
where the constant $\gamma$ is a dimensionless numerical parameter which should
be chosen of the order of magnitude 1 and $\vec \tau$ the tangential vector
pointing in the direction of maximum shear traction. We now have managed to
reduce the equations to a linear system, but its computational cost is still
high as the linear system is about three times as large as the system for the
fluid flow. Choosing the optimal step length parameter is therefore crucial to
obtain a viable scheme. After updating the slip distance for all faces
violating the Mohr-Coulomb criterion, we go back to solving
Equation~\eqref{berge_mini_2:eq:elastic_equation}, and iterate in this way
until Equation~\eqref{berge_mini_2:eq:frac_equation_mohr}
and~\eqref{berge_mini_2:eq:frac_equation_slip} is satisfied. Note that we might
update the slip distance for a fracture face several times per time step before
the excess shear stress is reduced to zero.

\section{Numerical Results}
\label{berge_mini_2:sec:simulation}

In this section, we will present simulations of two different reservoirs, one
with a single fracture, and the second with several fractures.

The purpose of the first example is twofold; we wish to validate that the slip
length in fact scales with shear modulus and the mesh size, and we will try to
find an optimal $\gamma$. Meshes of two different sizes are used, one fine and
one coarse, and two different shear modulus parameters $G$. We will vary
$\gamma$ and count the number of iterations per time step and look at the error
in approximating Equation~\ref{berge_mini_2:eq:frac_equation}. We stop the
iteration procedure when $T_e/G <1\text{e}{-5}$. The geometric domain is
described by a single circular fracture with a radius of 1500~m. Fluid is
injected at the center of the fracture at a constant rate of $1$~L/s.

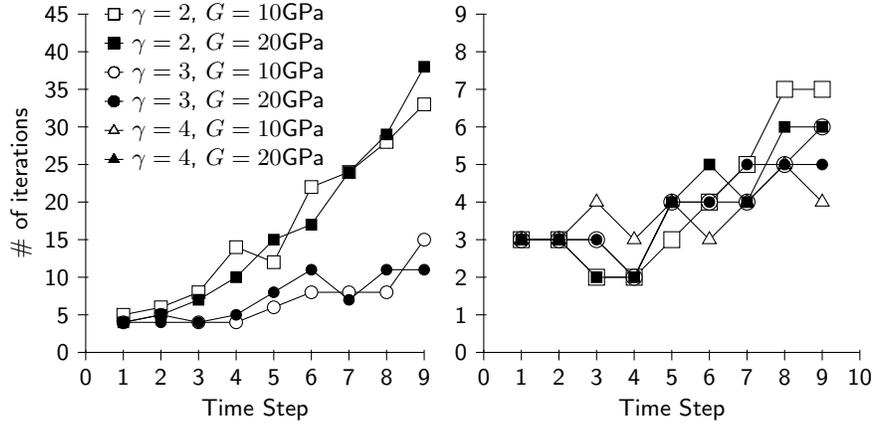
\begin{figure}
  \begin{tikzpicture}[y=.1cm, x=.5cm,font=\sffamily]
    % axis
    \draw (0,0) -- coordinate (x axis mid) (9,0); \draw (0,0) -- coordinate (y
    axis mid) (0,45);
    % ticks
    \foreach \x in {0,...,9} \draw (\x,1pt) -- (\x,-3pt) node[anchor=north]
    {\x}; \foreach \y in {0,5,...,45} \draw (1pt,\y) -- (-3pt,\y)
    node[anchor=east] {\y};
    % labels
    \node[below=0.5cm] at (x axis mid) {Time Step}; \node[rotate=90,
    above=0.6cm] at (y axis mid) {\# of iterations};

    % plots
    \draw plot[mark=square*, mark options={fill=white,scale=1.2} ] file
    {gamma_2_shear_10_f.data}; \draw plot[mark=*, mark
    options={fill=white,scale=1.2}] file {gamma_3_shear_10_f.data};

    \draw plot[mark=square*, mark options={fill=black} ] file
    {gamma_2_shear_20_f.data}; \draw plot[mark=*, mark
    options={fill=black}] file {gamma_3_shear_20_f.data};

    % legend
    \begin{scope}[shift={(.5,45)}]
      \draw (0,0) -- plot[mark=square*, mark options={fill=white}] (0.25,0) --
      (0.5,0) node[right]{$\gamma= 2$, $G= 10$GPa};
      \draw[yshift=-1\baselineskip] (0,0) -- plot[mark=square*, mark
      options={fill=black}] (0.25,0) -- (0.5,0) node[right]{$\gamma= 2$,
        $G= 20$GPa};

      \draw[yshift=-2\baselineskip] (0,0) -- plot[mark=*, mark
      options={fill=white}] (0.25,0) -- (0.5,0) node[right]{$\gamma= 3$,
        $G= 10$GPa}; \draw[yshift=-3\baselineskip] (0,0) -- plot[mark=*, mark
      options={fill=black}] (0.25,0) -- (0.5,0) node[right]{$\gamma= 3$,
        $G= 20$GPa}; \draw[yshift=-4\baselineskip] (0,0) --
      plot[mark=triangle*, mark options={fill=white}] (0.25,0) -- (0.5,0)
      node[right]{$\gamma= 4$, $G= 10$GPa}; \draw[yshift=-5\baselineskip] (0,0)
      -- plot[mark=triangle*, mark options={fill=black}] (0.25,0) -- (0.5,0)
      node[right]{$\gamma= 4$, $G= 20$GPa};
    \end{scope}
  \end{tikzpicture}
  \begin{tikzpicture}[y=.5cm, x=.5cm,font=\sffamily]
    % axis
    \draw (0,0) -- coordinate (x axis mid) (10,0); \draw (0,0) -- coordinate (y
    axis mid) (0,9);
    % ticks
    \foreach \x in {0,...,10} \draw (\x,1pt) -- (\x,-3pt) node[anchor=north]
    {\x}; \foreach \y in {0,...,9} \draw (1pt,\y) -- (-3pt,\y)
    node[anchor=east] {\y};
    % labels
    \node[below=0.5cm] at (x axis mid) {Time Step};
    %	\node[rotate=90, above=0.8cm] at (y axis mid) {\# of iterations};

    % plots
    \draw plot[mark=square*, mark options={fill=white,scale=1.5} ] file
    {gamma_2_shear_10.data}; \draw plot[mark=*, mark
    options={fill=white,scale=1.5}] file {gamma_3_shear_10.data}; \draw
    plot[mark=triangle*, mark options={fill=white,scale=1.5} ] file
    {gamma_4_shear_10.data};

    \draw plot[mark=square*, mark options={fill=black} ] file
    {gamma_2_shear_20.data}; \draw plot[mark=*, mark options={fill=black}]
    file {gamma_3_shear_20.data}; \draw plot[mark=triangle*, mark
    options={fill=black} ] file {gamma_4_shear_20.data};

  \end{tikzpicture}
  \caption{Number of iterations needed for each time step to find the slip
    distance $\vec d$. The left and right figures show the number of iteration
    needed on a fine grid (852 fracture cells) and a coarse grid (226 fracture
    cell) respectively.\label{berge_mini_2:fig:num_it} }
\end{figure}

The total number of iterations at each time step is shown in
Fig.~\ref{berge_mini_2:fig:num_it}. The number of iterations is relatively low
at the beginning of the injection, but increases with time.  At the first time
steps, only a few cells in the vicinity of the injection cells slip. As the
pressure front moves radially out from the center, more cells slip at each time
step. When several cells slip in the same region, the assumption we made when
updating the slip distance becomes more erroneous, and we need additional
iterations to reach the stop criterion. For the coarse grid, the slip is more
or less limited to one or two rows of cells around the pressure front, so the
iteration numbers stay low. For the fine grid many more cells slip, which
causes the number of iterations to increase drastically. Increasing the step
parameter $\gamma$ does as expected reduce the number of iterations. However,
as $\gamma$ increases, the overshoot of the slip distance can also increase
(meaning that the excess shear stress $T_e$ becomes negative). For the fine
grid, we did not include a simulation of $\gamma = 4$ as the scheme did not
converge for this value. The maximum overshoot error $\lvert T_e\rvert / G$
were on the order of magnitude $1\text e{-4}$ for $\gamma = 3,\ 4$ and
$1\text e{-5}$ for $\gamma = 2$. Typically, the error for only one or two cells
came close to the maximum value, while the error for the other cells were much
smaller.  \enlargethispage{\baselineskip}

\begin{figure}
  \centering \def\svgwidth{0.7 \textwidth}%
  \begingroup%
  \makeatletter%
  \providecommand\color[2][]{%
    \errmessage{(Inkscape) Color is used for the text in Inkscape, but the
      package 'color.sty' is not loaded}%
    \renewcommand\color[2][]{}%
  }%
  \providecommand\transparent[1]{%
    \errmessage{(Inkscape) Transparency is used (non-zero) for the text in
      Inkscape, but the package 'transparent.sty' is not loaded}%
    \renewcommand\transparent[1]{}%
  }%
  \providecommand\rotatebox[2]{#2}%
  \ifx\svgwidth\undefined%
  \setlength{\unitlength}{1621.63251953bp}%
  \ifx\svgscale\undefined%
  \relax%
  \else%
  \setlength{\unitlength}{\unitlength * \real{\svgscale}}%
  \fi%
  \else%
  \setlength{\unitlength}{\svgwidth}%
  \fi%
  \global\let\svgwidth\undefined%
  \global\let\svgscale\undefined%
  \makeatother%
  \begin{picture}(1,0.86189174)%
    \put(0,0){\includegraphics[width=\unitlength]{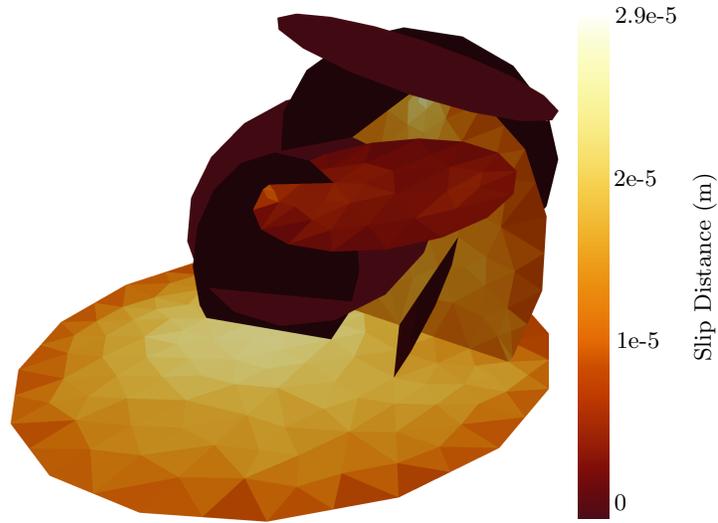}}%
    \put(1.0076493,0.0333521){\color[rgb]{0,0,0}\makebox(0,0)[lb]{\smash{0}}}%
    \put(1.00678211,0.56115277){\color[rgb]{0,0,0}\makebox(0,0)[lb]{\smash{2e-5}}}%
    \put(1.00877964,0.82505308){\color[rgb]{0,0,0}\makebox(0,0)[lb]{\smash{2.9e-5}}}%
    \put(1.16431083,0.42856235){\color[rgb]{0,0,0}\rotatebox{90}{\makebox(0,0)[b]{\smash{Slip
            Distance (m)}}}}%
    \put(1.01077175,0.29738252){\color[rgb]{0,0,0}\makebox(0,0)[lb]{\smash{1e-5}}}%
  \end{picture}%
  \endgroup%
  \caption{Total slip at equlibrium after shut-in of well for case 2. The well
    is located at the center of the biggest
    fracture.\label{berge_mini_2:fig:slip_equlibrium}}
\end{figure}

In the second case, we apply the algorithm to a small fracture network of 8
fractures. The size of the fractures varies from a few hundred meters up to
1500 meters. We inject water at the center of the largest fracture with a
constant rate of $30$ L/s for $30$
minutes. Fig.~\ref{berge_mini_2:fig:slip_equlibrium} depicts the total slip
distance at the end of the simulation. We register slip after the shut-in of
the well as the pressure migrates through the fracture network. At the end time
equilibrium has been reached, and we observe how the slip distance varies
significantly between different fractures. This is due to the fracture's
orientation relative to the background stress field, and how far they are
located from the injection well.

\section{Concluding Remarks}
\label{berge_mini_2:sec:conclusions}
This paper has presented a coupled model of compressible fluid flow, elastic
rock deformation and plastic fracture deformation. A simple iterative scheme to
handle the non-linear coupling between the rock and fracture deformation were
discussed, and we suggested an improvement to this scheme by showing how the
step length parameter should depend on the mesh size and shear modulus. The
numerical results support the choice of step length. 
\def\cprime{$'$}

\section*{Acknowledgments}
This work was partially funded by the Research Council of Norway, TheMSES
project, grant no. 250223, and travel support funded by Statoil, through the
Akademia agreement.

% 1. BibTeX version:
%
% If you want to choose this version uncomment the next two lines and insert
% the name of the BibTeX file.
%
%
% \bibliographystyle{vmams}
% \bibliography{ref} % use BibTeX file example.bib (BibTeX -> author.bbl)

\ifx\undefined\bysame \newcommand{\bysame}{\leavevmode\hbox
  to3em{\hrulefill}\,} \fi

\end{document}